\documentclass[aps,preprint,prd,showpacs,nofootinbib]{revtex4}

\usepackage{amsmath,amssymb}
\usepackage{graphicx,subfigure}
\usepackage{color,multirow}
\usepackage[colorlinks,linkcolor=magenta,anchorcolor=cyan,citecolor=blue,plainpages=false]{hyperref}

\hypersetup{colorlinks=true,
    breaklinks=true,
    pdfstartview=Fit,
    linkcolor=blue,
    citecolor=blue,
    urlcolor=blue}

\def\be{\begin{equation}}
    \def\ee{\end{equation}}
\def\ba{\begin{eqnarray}}
    \def\ea{\end{eqnarray}}

\begin{document}

\title{Impact of evolving dark energy on the search for primordial gravitational waves}

\author{Hao Wang$^{1,2} $\footnote{\href{wanghao187@mails.ucas.ac.cn}{wanghao187@mails.ucas.ac.cn}}}
\author{Gen Ye$^{3} $\footnote{\href{ye@lorentz.leidenuniv.nl}{ye@lorentz.leidenuniv.nl}}}
\author{Yun-Song Piao$^{1,2,4,5} $ \footnote{\href{yspiao@ucas.ac.cn}{yspiao@ucas.ac.cn}}}

    \affiliation{$^1$ School of Fundamental Physics and Mathematical
        Sciences, Hangzhou Institute for Advanced Study, UCAS, Hangzhou
        310024, China}

    \affiliation{$^2$ School of Physics Sciences, University of
        Chinese Academy of Sciences, Beijing 100049, China}

     \affiliation{$^3$ Leiden University, Instituut-Lorentz for Theoretical Physics, 2333CA, Leiden,
     Netherlands}

    \affiliation{$^4$ International Center for Theoretical Physics
        Asia-Pacific, Beijing/Hangzhou, China}

    \affiliation{$^5$ Institute of Theoretical Physics, Chinese
        Academy of Sciences, P.O. Box 2735, Beijing 100190, China}

    \begin{abstract}

Recent data seem to suggest a \textit{preference} for the evolving
dark energy (DE). However, if the case is actually so, and not
caused by unknown systematics in data, it might impact our
understanding about our Universe in an anomalous way due to the
shifts of some primary parameters. As an example, we present the
search for the primordial gravitational waves, based on the
evolving DE. The joint analysis of recent BICEP/Keck cosmic
microwave background (CMB) B-mode polarization data with Planck18
CMB, DESI baryon acoustic oscillations and PantheonPlus data shows
that the bestfit tensor-to-scalar ratio is $r_{0.05}\sim 0.01$,
and the lower bound of $r_{0.05}$ is $\sim 2\sigma$ non-zero.

    \end{abstract}

    \maketitle

\section{INTRODUCTION}

It is well-known that inflation is the current paradigm of early
universe
\cite{Guth:1980zm,Linde:1981mu,Albrecht:1982wi,Starobinsky:1980te},
which predicts nearly scale-invariant scalar perturbation
consistent with the cosmic microwave background (CMB)
observations, as well as the primordial gravitational waves (GWs).

The ultra-low-frequency primordial GWs at $f\sim
10^{-18}-10^{-16}$Hz, thought as the ``smoking gun" of inflation,
can source the B-mode polarization in the CMB
\cite{Seljak:1996ti,Kamionkowski:1996zd,Seljak:1996gy}, and search
for it is currently the most promising way to detect the
primordial GWs. Based on the standard $\Lambda$CDM model, using
Planck18 CMB, baryon acoustic oscillations (BAO) and BICEP/Keck18
CMB B-mode dataset the BICEP/Keck collaboration has reported the
upper bound on the tensor-to-scalar ratio, $r<0.036$ (95\% C.L.)
\cite{BICEP:2021xfz}, see also \cite{Tristram:2021tvh}. The models
with larger Hubble constant will have tighter upper bound, up to
$r<0.028$ (95\% C.L.) \cite{Ye:2022afu,Jiang:2023bsz}. Recently, a
joint analysis of NANOGrav 15-year (at $f\sim 10^{-10}-10^{-8}$Hz)
\cite{NANOGrav:2023gor} and BICEP/Keck18 dataset has showed
$r<0.039$ (95\% C.L.) \cite{Jiang:2023gfe}, see also
\cite{NANOGrav:2023hvm,Vagnozzi:2020gtf,Vagnozzi:2023lwo}.

The standard $\Lambda$CDM model is thought to be the most
successful model explaining most of cosmological observations, in
which $\Lambda$ is a positive cosmological constant and also call
the dark energy (DE) responsible for the accelerated expansion of
our current Universe. In the past decades, identifying the nature
of DE has been still an important challenge. Recently, using their
first year data the DESI collaboration
\cite{DESI:2024lzq,DESI:2024mwx,DESI:2024uvr} has found a $\gtrsim
3\sigma$ evidence for an evolving DE.
The relevant issues are being intensively investigated
\cite{DESI:2024aqx,DESI:2024kob,Wang:2024dka,Wang:2024pui,Yang:2024kdo,Yin:2024hba,Luongo:2024fww,Cortes:2024lgw,Carloni:2024zpl,Wang:2024hks,Wang:2024rjd,Colgain:2024xqj,Giare:2024smz,Escamilla-Rivera:2024sae,Park:2024jns,Shlivko:2024llw,Dinda:2024kjf,Seto:2024cgo,Bhattacharya:2024hep,Roy:2024kni,Wang:2024hwd,Notari:2024rti,Heckman:2024apk,Gialamas:2024lyw,Orchard:2024bve}

The result of DESI collaboration about an evolving DE has
astonished the scientific community, It has been also observed
that this anomalous result is caused primarily by the LRG samples
at low redshift \cite{Wang:2024pui,Chudaykin:2024gol,Liu:2024gfy}.
However, if the dataset actually preferred the evolving DE, it
might affect our understanding about our Universe in an unexpected
way, thus it is significant to explore the underlying impact of
evolving DE on the physics of very early Universe. As a step
towards this issue, we present the search for primordial GWs, and
find \be r_{0.05}=0.0153^{+0.0051+0.0188}_{-0.0133-0.0155}\quad
(68\%\,\, \mathrm{and}\,\, 95\%\,\, \mathrm{CL}.)\ee The upper
bound is consistent with that reported by the BICEP/Keck
collaboration \cite{BICEP:2021xfz}, but the lower bound is
$\sim$1.9$\sigma$ non-zero. The bestfit $r$ is $r_{0.05}\sim
0.01$, see also Fig.1.

\begin{figure*}
    \includegraphics[width=0.7\columnwidth]{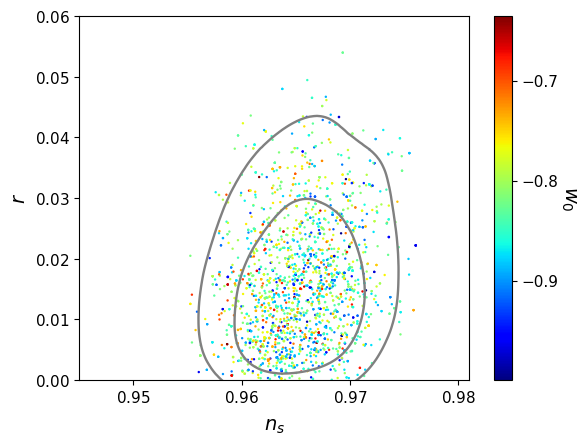}
    \caption{\label{rns} The $r-n_s$ contours for the
$w_0w_a$CDM model, fitting to Planck18+BK18+DESI+Pantheon Plus
datasets, where $n_s$ is the spectral index of primordial scalar
perturbation.}
\end{figure*}

\section{DATA AND METHODOLOGY}
Recent DESI data is based on samples of bright galaxies, LRGs, ELGs, quasars and Ly$\alpha$
Forest at the redshift region $0.1<z<4.2$
\cite{DESI:2024mwx,DESI:2024lzq,DESI:2024uvr}. Here, we will use
their measurements for the comoving distances $D_M(z)/r_d$ and
$D_H(z)/r_d$, where
    \begin{equation}\label{DMDH}
        D_M(z)\equiv\int_{0}^{z}{cdz'\over H(z')},\quad D_H(z)\equiv {c\over
        H(z)},
    \end{equation}
and $r_d=\int_{z_d}^{\infty}{c_s(z)\over H(z)}$ is the sound
horizon with $z_d\simeq1060$ at the baryon drag epoch and $c_s$
the speed of sound, as well as the angle-averaged quantity
$D_V/r_d$, where $D_V(z)\equiv\left(zD_M(z)^2D_H(z)\right)^{1/3}$.
    \begin{table}[htbp]
        \centering
        \begin{tabular}{c|c|cc|c}
            tracer&$z_{eff}$&$D_M/r_d$&$D_H/r_d$&$D_V/r_d$\\
            \hline
            BGS&0.30&-&-&$7.93\pm0.15$\\
            LRG&0.51&$13.62\pm0.25$&$20.98\pm0.61$&-\\
            LRG&0.71&$16.85\pm0.32$&$20.08\pm0.60$&-\\
            LRG+ELG&0.93&$21.71\pm0.28$&$17.88\pm0.35$&-\\
            ELG&1.32&$27.79\pm0.69$&$13.82\pm0.42$&-\\
            QSO&1.49&-&-&$26.07\pm0.67$\\
            Lya QSO&2.33&$39.71\pm0.94$&$8.52\pm0.17$&-\\
        \end{tabular}
\caption{\label{DESI}Statistics for the DESI samples of the DESI
DR1 BAO measurements used in this paper.}
    \end{table}

In addition, we also use \textbf{Planck 2018 CMB} dataset (low-l
and high-l TT, TE, EE spectra, and reconstructed CMB lensing
spectrum \cite{Planck:2018vyg,Planck:2019nip,Planck:2018lbu}),
\textbf{Pantheon+} data (consisting of 1701 light curves of 1550
spectroscopically confirmed Type Ia SN coming from 18 different
surveys \cite{Scolnic:2021amr}), and the CMB B-mode polarization
data \textbf{BK18} \cite{BICEP:2021xfz}. Throughout this paper, we
work with \textbf{Planck18+BK18+DESI+Pantheon Plus} datasets.

We modified the MontePython-3.6 sampler
\cite{Audren:2012wb,Brinckmann:2018cvx} and CLASS codes
\cite{Lesgourgues:2011re,Blas:2011rf} to perform our MCMC
analysis. Here, to observe the impact of evolving DE on $r$ in
light of DESI and BK18, we consider the $w_0w_a$CDM model with the
equation of state of DE $w_\mathrm{DE}=w_0+w_a{z\over 1+z}$
\cite{Chevallier:2000qy,Linder:2002et}. The pivot scale
$k_\mathrm{pivot}$ of $r$ is set to 0.05(Mpc)$^{-1}$. In
Table.\ref{prior} we list the flat priors used for our MCMC
\textit{parameters} \{$\omega_b$, $\omega_{cdm}$, $H_0$,
$\ln10^{10}A_s$, $n_s$, $\tau_{reio}$, $w_0$, $w_a$, $r$\}. We
adopt a Gelman-Rubin convergence criterion with a threshold
$R-1<0.01$.

In order to see to what extent the results on $r$ are
model-dependent, we also consider the $w_0w_a$CDM model with a
negative cosmological constant (nCC), which is hinted by DESI
according to recent Refs.\cite{Wang:2024hwd,Notari:2024rti}, see
\cite{Dutta:2018vmq,Visinelli:2019qqu,Ruchika:2020avj,Calderon:2020hoc,Sen:2021wld,Malekjani:2023ple}
for earlier works \footnote{The negative CC, which corresponds to
the anti-de Sitter (AdS) vacuum, might contribute to settle the
Hubble tension, see e.g. AdS-EDE
\cite{Ye:2020btb,Ye:2020oix,Jiang:2021bab,Ye:2021iwa,Wang:2022jpo}
and the switch-sign CC
\cite{Akarsu:2019hmw,Akarsu:2021fol,Akarsu:2022typ,Akarsu:2023mfb,Paraskevas:2024ytz,Anchordoqui:2023woo,Toda:2024ncp,Yadav:2024duq}.}.
We adapt a flat prior for the density parameter of nCC
$\Omega_L\in[-2,0.6]$ additionally. The density parameter
$\Omega_x$ of evolving $w_0w_a$ DE satisfies
$\Omega_\mathrm{DE}=\Omega_x+\Omega_L\simeq0.7$.

    \begin{table}[htbp]
    \centering
    \begin{tabular}{cc}
        \hline
        Parameters&Prior\\
        \hline
        $100\omega_b$&[None, None]\\
        $\omega_{cdm}$&[None, None]\\
        $H_0$&[65, 80]\\
        $\ln10^{10}A_s$&[None, None]\\
        $n_s$&[None,None]\\
        $\tau_{reio}$&[0.004, None]\\
        \hline
        $w_0$&[-2, 0.34]\\
        $w_a$&[-3,2]\\
        \hline
        $r$&[0,0.5]\\
        \hline
    \end{tabular}
\caption{\label{prior} The priors of primary parameters we adopt
in MCMC analysis.}
 \end{table}

\section{Results and Discussion}

\begin{figure*}
    \includegraphics[width=0.8\columnwidth]{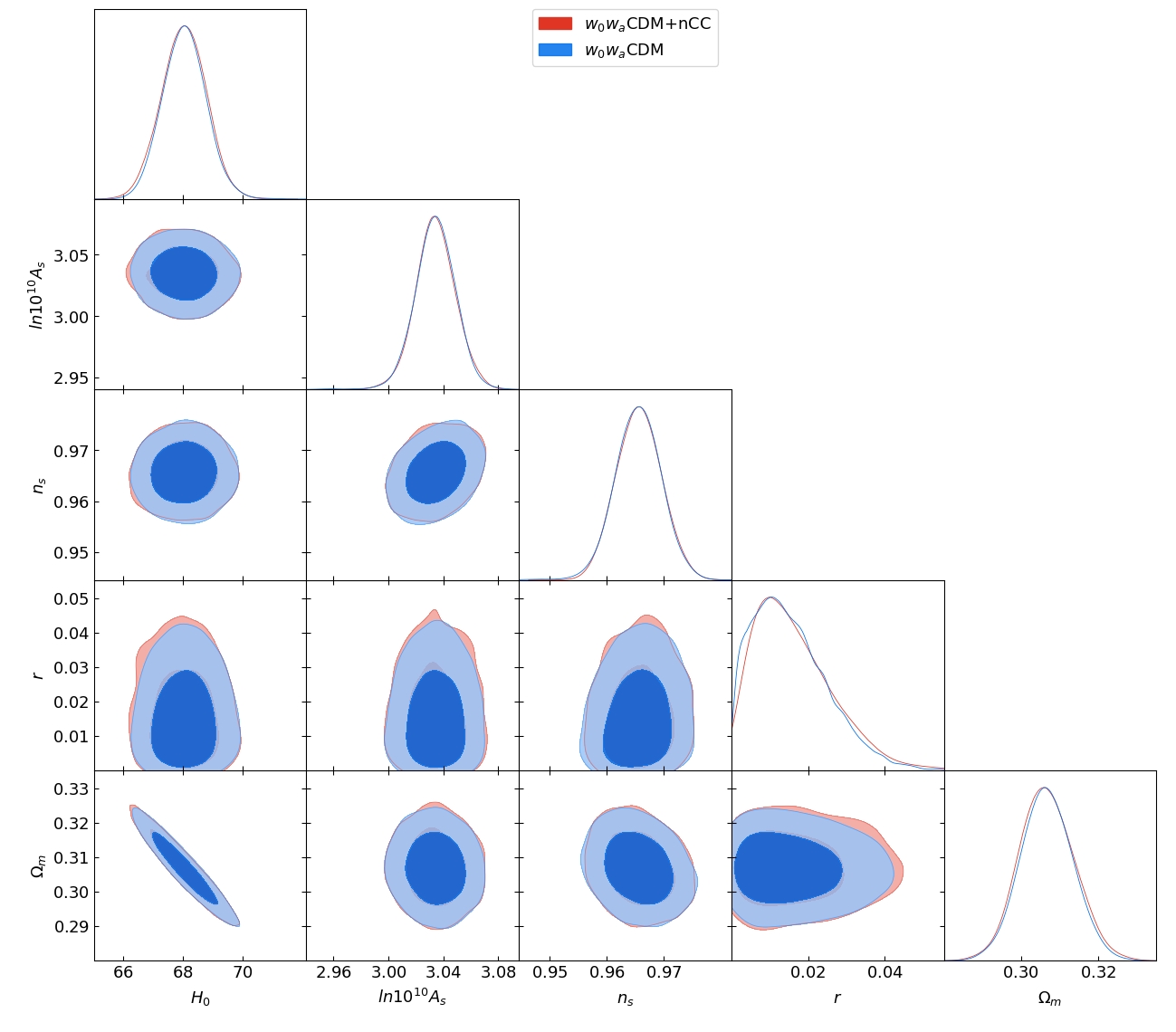}
    \caption{\label{MC} 2D contours of the
        primary parameters at 68\% and 95\% CL for the
evolving DE, fitting to Planck18+BK18+DESI+Pantheon Plus
datasets.}
\end{figure*}

\begin{figure}
    \includegraphics[width=0.8\columnwidth]{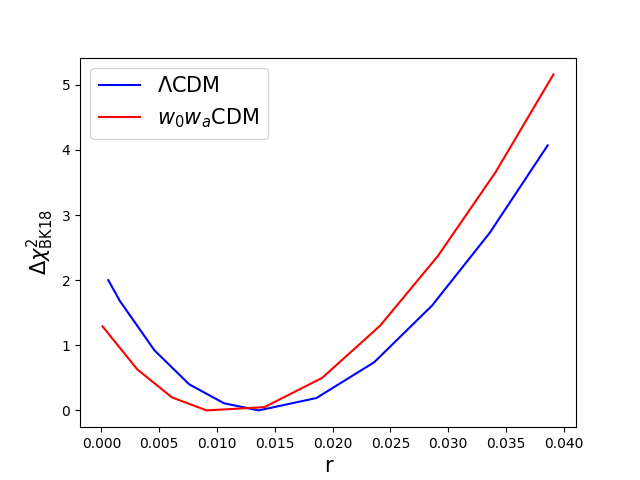}
\caption{\label{chi2}Response of total BK18 $\chi^2$ to the
variation in $r$. $\chi^2$ at each point is calculated by varying
$r$ with all other parameters fixed to their bestfit values. The
y-axis plots $\Delta\chi^2=\chi^2-\chi^2_\mathrm{bestfit}$ for the
BK18 datasets (here the range of $\Delta \chi^2=1$ is
smaller than the actual 1$\sigma$). As a comparison, we also plot
that for the $\Lambda$CDM model, see Appendix.}
\end{figure}

In Table.\ref{MCtable}, we present our MCMC results. The $1\sigma$
and $2\sigma$ marginalized posterior distributions of the
corresponding parameters are showed in Fig.\ref{MC}. As expected,
an evolving DE is preferred with $w_0=-0.832\pm0.070$ and
$w_a=-0.667\pm0.333$, consistent with those of DESI collaboration
\cite{DESI:2024mwx}. The result of tensor-to-scalar ratio $r$ is
$r=0.0153^{+0.0051}_{-0.0133}$ at the 1$\sigma$(68\%) CL. and
$r=0.0152^{+0.0188}_{-0.0155}$ at the 2$\sigma$ (95\%) CL., its
lower bound is $\sim$1.9$\sigma$ non-zero. The response of total
BK18 $\chi^2$ to the variation in $r$ is showed in Fig.\ref{chi2}.
We can clearly see that the bestfit value is $r=0.009$.

The $w_0w_a$CDM+nCC model shows similar but slightly smaller
result, $r=0.007$. And the bestfit CC is $\Omega_L\sim -0.75<0$.
It is noteworthy that the inclusion of a negative CC makes $w_0$
and $w_a$ closer to $w_0=-1$ and $w_a=0$, in particular,
$w_0+w_a\geqslant -1$ can be $\sim 1\sigma$ consistent, see also
\cite{Wang:2024hwd}. In corresponding scalar field model our
result suggest that the AdS vacuum might exist and at low redshift
the scalar field will roll towards it.

    \begin{table*}[htbp]
    \centering
    \begin{tabular}{c|c|c}
        \hline
        Parameters&$w_0w_a$CDM&$w_0w_a$CDM+nCC\\
        \hline
        $100\omega_b$&2.231(2.238)$\pm$0.015&2.233(2.232)$\pm$0.014\\
        $\omega_{cdm}$&0.120(0.120)$\pm$0.001&0.120(0.120)$\pm$0.001\\
        $H_0$&68.07(68.48)$\pm$0.83&68.03(67.95)$\pm$0.75\\
        $\ln10^{10}A_s$&3.034(3.041)$\pm$0.015&3.034(3.046)$\pm$0.015\\
        $n_s$&0.965(0.966)$\pm$0.004&0.966(0.967)$\pm$0.004\\
        $\tau_{reio}$&0.052(0.054)$\pm$0.008&0.052(0.058)$\pm$0.008\\
        \hline
        $w_0$&-0.832(-0.857)$\pm$0.070&-0.902(-0.956)$\pm$0.064\\
        $w_a$&-0.667(-0.599)$\pm$0.333&-0.544(-0.237)$\pm$0.707\\
        $\Omega_L$&-&-1.08(-0.749)$\pm$0.65\\
        $r$&0.0153(0.0091)$^{+0.0051}_{-0.0133}$&0.0160(0.0072)$^{+0.0066}_{-0.0127}$\\
        \hline
        $\Omega_m$&0.307(0.303)$\pm$0.008&0.307(0.308)$\pm$0.007\\
        $S_8$&0.833(0.834)$\pm$0.012&0.833(0.842)$\pm$0.012\\
        \hline
        $\chi^2_\mathrm{CMB}$&2771.07&2771.09\\
        $\chi^2_\mathrm{DESI}$&14.01&13.78\\
        $\chi^2_\mathrm{BK18}$&537.52&537.31\\
        $\chi^2_\mathrm{Pantheon+}$&1409.14&1409.03\\
        \hline
        $\chi^2_\mathrm{tot}$&4731.22&4730.71\\
        \hline
    \end{tabular}
\caption{\label{MCtable} Mean (bestfit) values and 1$\sigma$
regions of the parameters of models, fitting to
Planck18+BK18+DESI+Pantheon Plus datasets.}
\end{table*}


As pointed out in Ref.\cite{Ye:2022afu}, the constraint on $r$ is
related to the amplitude of the scalar perturbation $A_s$. A
larger $A_s$ contributes to larger $C_{l,\mathrm{lensing}}^{BB}$,
and smaller $C_{l,\mathrm{tensor}}^{BB}$ since
$C_{l,\mathrm{total}}^{BB}=C_{l,\mathrm{lensing}}^{BB}+C_{l,\mathrm{tensor}}^{BB}$
is set by BK18 dataset, and thus a smaller $r$. This negative
relation between the bestfit values of $A_s$ and $r$ can be seen
in Table.\ref{MCtable}. However, the dataset preferring the
evolving DE does not simply shift the posteriors of $A_s$. Both
$w_0w_a$CDM and $w_0w_a$CDM+nCC models also accommodate larger
$H_0$ and smaller $\Omega_m$, compared with $\Lambda$CDM with
pre-DESI data (e.g.\cite{Ye:2022afu}), and thus the posterior of
$A_s$, also $r$, is enlarged to both larger and smaller sides.
This also explains why the posteriors of $r$ in both models are
similar but their bestfit values slightly differ.

\begin{figure*}
    \includegraphics[width=0.8\columnwidth]{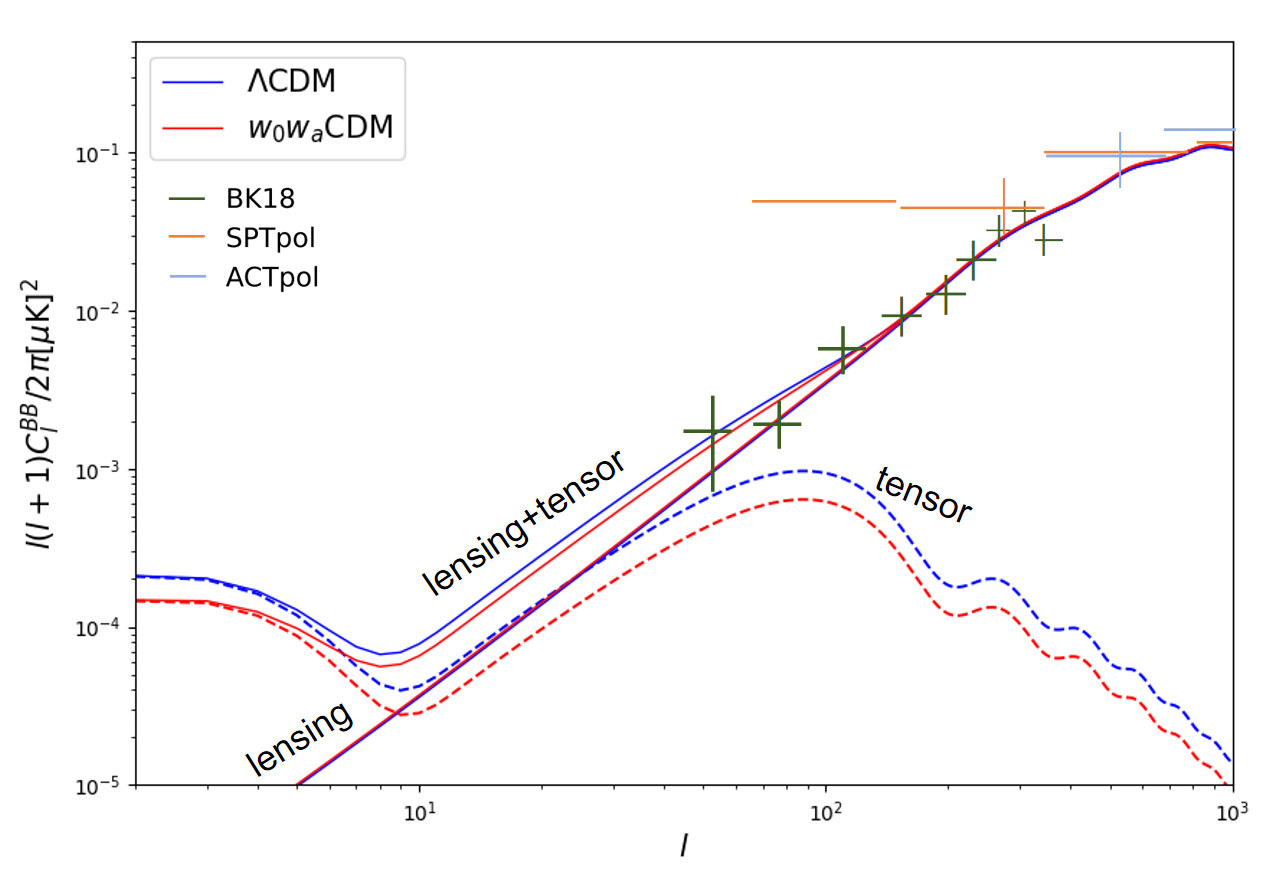}
\caption{\label{EE} $C_{l,\mathrm{total}}^{BB}$,
$C_{l,\mathrm{lensing}}^{BB}$, $C_{l,\mathrm{tensor}}^{BB}$ for
the bestfit values of the $w_0w_a$CDM and $\Lambda$CDM models.
Points with error bars are binned BK18 \cite{BICEP:2021xfz}, SPT
\cite{SPT:2019nip} and ACT \cite{ACT:2020frw} data points.}
\end{figure*}

It is also interesting to compare our result with that of the
$\Lambda$CDM model. In Appendix, we also confront the $\Lambda$CDM
model with the Planck18+BK18+DESI+Pantheon Plus datasets, though
it is not the bestfit model of this dataset (its $\chi^2$ is
larger,
$\chi^2_{\Lambda\mathrm{CDM}}-\chi^2_{w_0w_a\mathrm{CDM}}=7.47$).
The evolving DE has a significant impact on the cosmological
evolution at low redshift $z\lesssim0.5$, which affects the
perturbation spectrum $P(k,z)$, and thus needs to adjust the
original parameters to maintain consistency with observations. As
plotted in Fig.\ref{EE}, $C_{l,\mathrm{lensing}}^{BB}$ of
$w_0w_a$CDM is slightly larger than that of $\Lambda$CDM, since
$\delta\ln10^{10}A_s\sim+0.005$, and thus
$C_{l,\mathrm{tensor}}^{BB}$ is smaller, $r=0.009$, to keep
$C_{l,\mathrm{total}}^{BB}$ nearly unchanged.

In certain sense, we are testing the potential impact of evolving
DE on the search for the primordial GWs. In the light of our work,
if recent datasets actually preferred the evolving DE (which is
not due to some unknown systematics in data), our approach to $r$
and its extension might be worth further investigating, which
would bring the unexpected insights into our primordial Universe,
in particular the inflation models.

\section*{Acknowledgments}

YSP is supported by National Key Research and Development Program
of China (Grant No. 2021YFC2203004), NSFC (Grant No.12075246), and
the Fundamental Research Funds for the Central Universities. GY is
supported by NWO and the Dutch Ministry of Education, Culture and
Science (OCW) (Grant VI.Vidi.192.069).

\appendix

\section{MCMC results of $\Lambda$CDM}\label{appendix}

Here, we confront the $\Lambda$CDM model with the
Planck18+BK18+DESI+Pantheon Plus datasets. The results are
presented in Table.\ref{MCtable-L}, see also recent results using
Planck+BK18+DESI dataset \cite{Wang:2024hks}.

\begin{table*}[htbp]
    \centering
    \begin{tabular}{c|c}
        \hline
        Parameters&$\Lambda$CDM\\
        \hline
        $100\omega_b$&2.235(2.238)$\pm$0.014\\
        $\omega_{cdm}$&0.119(0.119)$\pm$0.001\\
        $H_0$&68.17(68.43)$\pm$0.41\\
        $\ln10^{10}A_s$&3.041(3.035)$\pm$0.014\\
        $n_s$&0.967(0.966)$\pm$0.004\\
        $\tau_{reio}$&0.055(0.051)$\pm$0.007\\
        \hline
        $r$&0.0152(0.0136)$^{+0.0052+0.0184}_{-0.0130-0.0152}$\\
        \hline
        $\Omega_m$&0.305(0.301)$\pm$0.005\\
        $S_8$&0.827(0.818)$\pm$0.010\\
        \hline
        $\chi^2_\mathrm{CMB}$&2772.47\\
        $\chi^2_\mathrm{DESI}$&16.35\\
        $\chi^2_\mathrm{BK18}$&536.54\\
        $\chi^2_\mathrm{Pantheon+}$&1413.83\\
        \hline
        $\chi^2_\mathrm{tot}$&4739.19\\
        \hline
    \end{tabular}
\caption{\label{MCtable-L} Mean (bestfit) values and 1$\sigma$
regions of the parameters of the $\Lambda$CDM model, fitting to
Planck18+BK18+DESI+Pantheon Plus datasets.}
\end{table*}

\end{document}